\documentclass[12pt]{article}
\usepackage{amssymb}
\usepackage{amsmath}
\usepackage{amsthm}
\usepackage{graphicx}

\theoremstyle{definition}

\large

\title{\bf{The epic journey from Kepler's laws to Newton's law of universal
       gravitation revisited}}

\author{Hai-Chau Chang\\ Department of Mathematics\\
        National Taiwan University\\ Taiper, Taiwan\\ \and Wu-Yi Hsiang\\
        Department of Mathematics\\ University of California\\ Berkeley, CA 94720, U. S. A.}

\begin{document}

\maketitle

\clearpage

\section{Introduction}

Historically, three outstanding achievements in the 17th century
heralded the grand opening of  modern science, namely, the
discovery of Kepler's laws on planet-motions; the study of
gravitation force and the formulation of basic principles of
mechanics by Galileo, and the mathematical analysis of the above
two that enabled  Newton to achieve the grand synthesis that
naturally led to his far-reaching generalization: \emph{The Law
of Universal Gravitation}. The publication of \emph{Philosophiae
Naturalis Principia Mathematica}  in 1687 not only created modern
science, but also started a profound revolution on our
understanding of the universe (i.e. the civilization of rational
mind).\\
In summary, the major results of the mathematical
analysis in Newton's Principia can be stated  as the following
four theorems, namely\newline \ \newline \emph{Theorem A}:
Kepler's second law holds (i.e. $\displaystyle \frac{dA}{dt }$ =
constant) if and only if  the acceleration (resp. force) is
pointing toward the center (i.e. centripetal).\\

\emph{Theorem I}: The Kepler's first and second laws imply that
the acceleration vector $\mathbf{a}$ is  pointing toward the sun
and with its magnitude equal to

\begin{equation*}
\frac{\pi^2}{2} \frac{(2a)^3}{T^2}
\frac{1}{r^2},
\end{equation*}

namely
\begin{equation*}
\mathbf{a} = \frac{\pi^2}{2}
\frac{(2a)^3}{T^2} \frac{1}{r^2}
\left( \begin{array}{l}
-\cos\theta \\ -\sin\theta%
\end{array} \right)
\tag{1}
\end{equation*}

\emph{Theorem II}: (The uniqueness theorem and the converse of
Theorem I) Suppose that the acceleration  vector is centripetal
and with its magnitude inversely proportional to the square of
distance, namely
\begin{equation*}
\mathbf{a} =
\frac{K}{r^2}\tag{2}
\left( \begin{array}{l}
-\cos\theta \\
-\sin\theta%
\end{array} \right)
\end{equation*}

Then, the motion satisfies Kepler's secord law and its orbit is a
conic section.\\
\emph{Theorem III}: The gravitation force of a
thin spherical shell with uniform (area-wise) density  exerting
on an outside particle P is equal to
\begin{equation*}
G\frac{Mm}{\overline{OP}^2}
\tag{3}
\end{equation*}

where $M$ (resp. $m$) is the total mass of the spherical shell
(resp. the mass of the particle at $P$) and  $\overline{OP}$ is
the distance between the center $O$ and $P$. [The following
Theorem $III^{\prime }$ is an immediate  corollary of Theorem
III.]\\
\emph{Theorem $III^{\prime }$}: Let $\Sigma_1$ and
$\Sigma_2$ be a pair of spherical bodies with radially uniform
densities (i.e. each of them can be decomposed into the union of
thin spherical shells of Theorem III).  Then the magnitude of the
(total) gravitation force between them is equal to

\begin{equation*}
G\frac{M_1 M_2}{\overline{O_1 O_2}^2}
\tag{4}
\end{equation*}

where $M_1$ (resp. $M_2$) are the total masses of $\Sigma_1$
(resp. $\Sigma_2 $) and $O_1$ (resp. $O_2$) are their centers.\\

However, the mathematical  analysis (mainly geometrical) that
Newton gave in Principia are rather difficult to understand,
although most steps are quite elementary. Currently (i.e. Fall of
2007), the authors are giving a course at National Taiwan
University, Taipei  entitled "\emph{selected topics on
mathematics and civilization}." This paper is a outcome of our
preparation of  lectures on this topic. In order to make such a
tour of revisiting Newton's epic journey enjoyable we tried to
provide alternative proofs of Theorems I, II, III that are
elementary, simple and with clean-cut ideas (cf. \S 3 and \S\ 4).
We hope the new proofs  of \S 3 and \S 4 will make the
\emph{revisiting of this epic journey} also understandable,
enjoyable, thus inspiring for earnest young students.

\section{Kepler's laws of planet-motions}

\subsection{Some remarks on the historical background of "pre-Kepler"
astronomy}

\begin{enumerate} \item[(1)] The fascinating puzzle of planet
motions:\\
Ever since remote ancient times, most civilizations
noticed the strange behaviors of five prominent stars, each of
them, wandering among the background of all the other "fixed"
stars, each in their unique patterns and with their individual
periods. They are called planets (i.e. wanderers) in Greek time,
and nowadays called Venus, Mercury, Mars, Jupiter and Saturn. In
ancient astronomy of many prominent civilizations, the study of
planet motions was naturally the central topic but it remained to
be a \emph{fascinating puzzle} up until the discovery of
Kepler's laws of planet-motions [K-2,K-3].

\item[(2)] Ptolemy and Copernicus:\\

Among various models on planet-motions of the pre-Kepler era
which enable us to provide more or less self-consistent
explainations of astronomical observations on planet-motions, the
Ptolemy model [Pt] and the Copernicus model [Co] are certainly
the most outstanding two. We shall only mention here the
following two points on their main features, namely.

\begin{enumerate}
\item[(i)] The Ptolemy model puts the earth at the center (i.e. geocentric),
while the Copernicus model puts the  sun at the center (i.e. heliocentric.)

\item[(ii)] Both models use the method of epicycles to achieve a
reasonably adequate fitting to astronomical  observations, which
were not that accurate anyway.
\end{enumerate}

\item[(3)] Tycho de Brahe:\\

Tycho de Brahe (1546-1601) was a Danish nobleman who was destined
to devote his entire life to astronomical observation. His
astronomic interest was inspired by the solar eclipse of 1560
(i.e. by the predictability of astronomical events); while the
occurrence of a conjunction of Jupiter and Saturn in 1563 led him
to realize the \emph{lack of accurate astronomical data}, which
further inspired him to upgrade the accuracy of his astronomical
intruments and observations. Anyhow, his striving for accuracy
made him well-prepared for the big event of the discovery of a
Nova on November 11, 1572; and his book "De nova stella" made him
a leading astronomer of the entire Europe, a pride of the Kingdom
of Denmark, and earned him the patronage of Denmark's king,
Frederick II. Frederick gave him the island of Hveen, on which he
built the observatory Uraniborg with the financial support of the
king and carried out \emph{nightly} observations for more than
twenty years, thus accumulating a treasure of astronomical data
on planet motions that the Kepler's monumental achievement was
based upon. \end{enumerate}

\subsection{Johannes Kepler (1571-1630) and Kepler's three laws of
planet-motions}

\begin{enumerate} \item[(1)] Kepler was one of the many children
of a poor family. Young Kepler won a sequence of scholarships
that enabled him to attend the University of T\"{u}bingen, where
he learned the Copernicus system from Michael M\"{a}stlin and
became a firm believer of the heliocentric theory. He was
preparing himself for a career as a Lutheran minister, however,
fate intervened to change his destiny. The sudden death of the
mathematics teacher of a high school at Gratz and the
recommendation of T\"{u}bingen Faculty for the substitute of such
a post teaching both mathematics and astronomy, thus starting
Kepler's life-time pursuit in astronomy. Of course, we should
mention another major intervention of fate that inspired him to
embark his life-long journey in the search of laws of
planet-motions.

\item[(2)] Mysterium Cosmographicum (1596):\\

In the Copernicus system, there are altogether six planets
revolving around the sun. To the pious young Kepler, such a
system is a perfect creation of God, the fact that there are
\emph{exactly six} planets (although it is not the case nowadays)
must have its profound reason. Anyhow, this underlying "profound
reason" was one of the mysteries of the universe that the young
teacher was earnestly searching for. According to Kepler himself,
a wonderful revelation occurred to him on July $19^{\mbox{th}}$
of 1595, namely, the reason must be that there are exactly five
Platonic solids (i.e. regular polyhedra) and each of them is
placed between the six "orbital spheres" such that it is the
inscribing (resp. circumscribing) polyhedra of one of the six
"orbital spheres". Thus, such a wonderful geometric structure not
only explains why there are exactly six planets, but it also
determines the ratios among the radii of the six orbital spheres.
This is the origin of Kepler's first book and he devoted his
entire life to study planet motions in order to verify his "wild
conjecture."

\item[(3)] Astronomia Nova (1609):\\

Kepler, of course, sent a copy of the above book to Tycho de
Brahe, and most likely, such a master of astronomy would dismiss
such a "wild conjecture" merely as a youthful fantasy. However,
he was impressed by the keen intelligence and bold originality of
this young astronomer. By the time of 1600, Tycho de Brahe needed
the mathematical talent of young Kepler to "understand" his
life-time astronomical observations, while Kepler needed the
access to Tycho's treasure of astronomical data to verify his
grand "mystery of universe". On Jan $1^{\mbox{st}}$ of 1600,
Kepler set out to join Tycho de Brahe in Prague to be his
assistant, up until the death of Tycho de Brahe in October of
1601. Subsequently, Kepler succeeded Tycho de Brahe to be the
imperial mathematician and got hold of the superb Tychonic data.
The eighteen months of conjunction between the two outstanding
astronomers was actually a personality mismatch, but it
miraculously accomplished one of the greatest "relay" in the
history of sciences. It took many years of Kepler's superhuman
endeavors and superb mathematical talent, only after many
setbacks, twists and turns and with tremendous perseverance and
ingenuity, he finally discovered the first law and the second law
on the motion of Mars [K-2], namely.\\

\emph{The first law}: Mars moves on an elliptical orbit with the
sun situated at one of its foci.\\
\emph{The second law}: The area sweeping across by the interval joining the Mars toward the
sun per unit time is a constant, as indicated in Figure 1.\\

\begin{center}
 \includegraphics[scale=0.5]{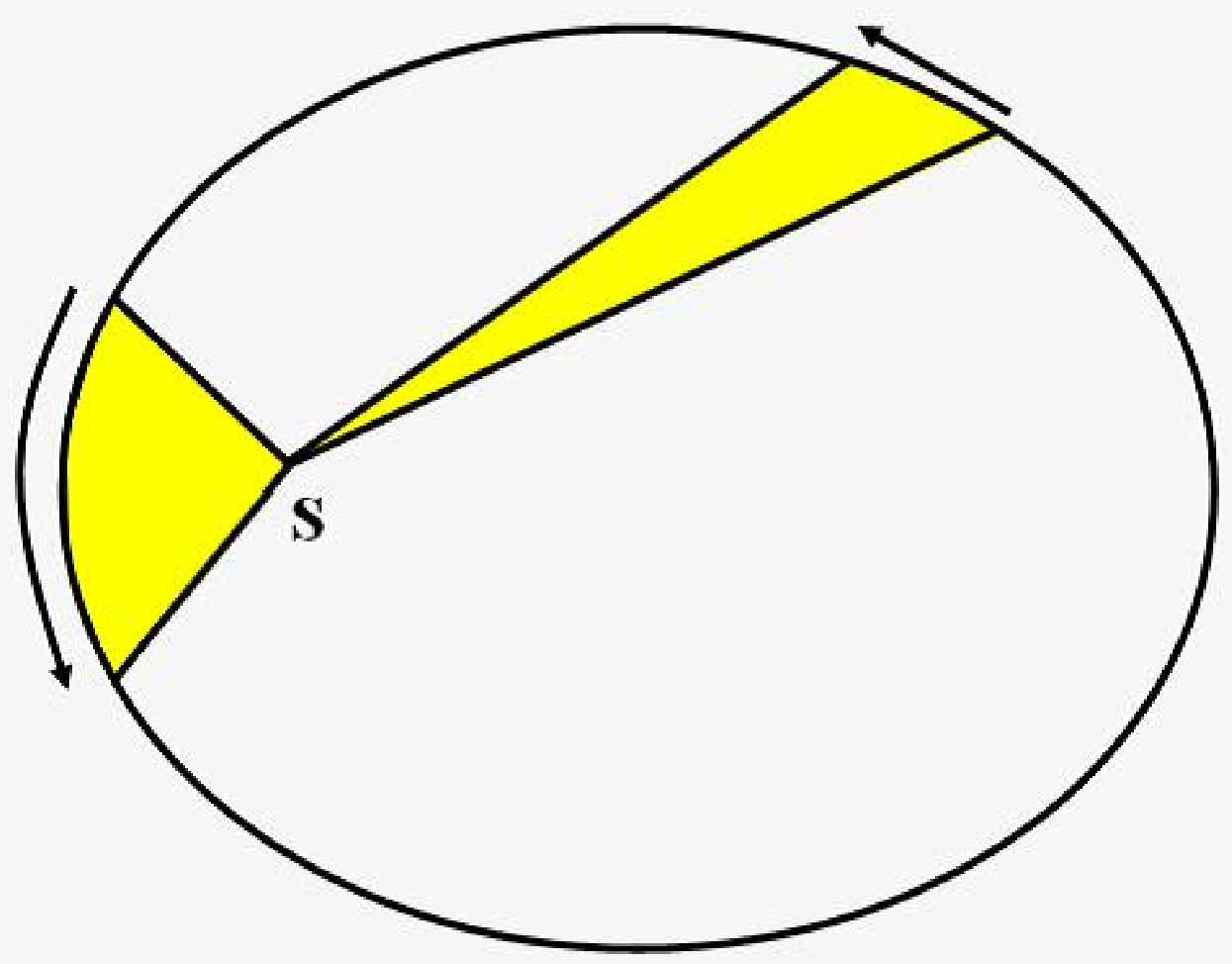}\\
 Figure 1
\end{center}

The above two empirical laws, based upon in depth mathematical analysis of
the observation data of Tycho de Brahe, heralded the grand opening of modern
astronomy.

\item[(4)] Harmonica Mundi (1619):\\

It took Kepler another decade of hard work to verify that the
same kind of first law and second law also hold for the other
five planets, and moreover, he discovered the following
remarkable third law, namely\\

The third law: The ratio between the cube of the major axis and
the square of period, i.e. $\displaystyle \frac{(2a)^3}{T^2}$,
remains the same for all the six planets. This was, indeed, a
wonderful reward for Kepler's life-long search of a kind of
harmony among planetary orbits. His youthful fantasy was somehow
vindicated. \end{enumerate}

\section{On the mathematical analysis of Kepler's laws}

Let us begin with the second law on planet motions, whose discovery actually
precedes that of the first law [K-2], and then proceed to the mathematical
analysis of both the first and the second laws jointly. As it has been discussed
in $\S 1$, such a journey was the monumental contribution of
Newton (cf. Principia). However, the proof of the latter given in Principia
is quite involved and rather difficult to grasp (i.e. understand) his
original insight that led to such a proof. In this section, we shall present
two simple and straightforward alternatives of the latter whose underlying
geometric ideas are rather clear.

\subsection{Mathematical analysis of Kepler's second law}

This is the easy part of the journey but it is a "good beginning" of basic
importance. Historically, the discovery of the second law (cf. [K-2]) not
only preceded that of the first law, but it also provided the crucial and
advantageous stepping-stone that eventually led him to the discovery of the
first law. Moreover, in essentially the same way, the understanding of the
mathematical as well as the physical meaning of Kepler's second law was also
the good beginning for Newton's journey. Anyhow, this step is very simple,
straightforward but very important.

\begin{enumerate}
\item[(1)] First of all, the second law is, by itself, \emph{local} in
nature. Let $(r,\theta)$ be the polar  coordinates of $P$ (i.e., the
position of the planet) with the position of the sun as the origin. Then,
the second  law of Kepler simply asserts that
\begin{equation*}
\frac{dA}{dt}=\frac{1}{2}r^{2}\omega, \: \: \: \omega=\frac{d\theta}{dt}\smallskip %
\mbox{(angular velocity)}  \tag{5}
\end{equation*}
is equal to a constant k, which is in fact equal to the total area divided
by the period T , namely.
\begin{equation*}
\mbox{total area}=\int^{T}_{0} dA=\int^{T}_{0}kdt=k\cdot T \tag{6}
\end{equation*}
Hence, in conjunction with the first law, one has the following powerful
simple equation
\begin{equation*}
r^2 \omega=\frac{2\pi ab}{T} (=2k) \tag{7}
\end{equation*}

\item[(2)] Let $\mathbf{v}$ (resp. $\mathbf{n}$) be the velocity (resp.
upward unit normal) vector. Then, as indicated in Figure 2.
\begin{equation*}
\overrightarrow{OP}\times \mathbf{v}=2k\mathbf{n}\tag{8}
\end{equation*}%
and hence
\begin{equation*}
\frac{d}{dt}(\overrightarrow{OP}\times \mathbf{v})=\mathbf{v}\times \mathbf{v%
}+\overrightarrow{OP}\times \mathbf{a}=0, \: \: \: \left( \mathbf{a}=\frac{d}{dt}%
\mathbf{v}\right) \tag{9}
\end{equation*}%
which implies that the acceleration vector $\mathbf{a}$ is \emph{collinear}
with $\overrightarrow{OP}$, namely, the force, m$\mathbf{a}$, is \emph{%
centripetal}.%

\begin{center}
 \includegraphics[scale=0.5]{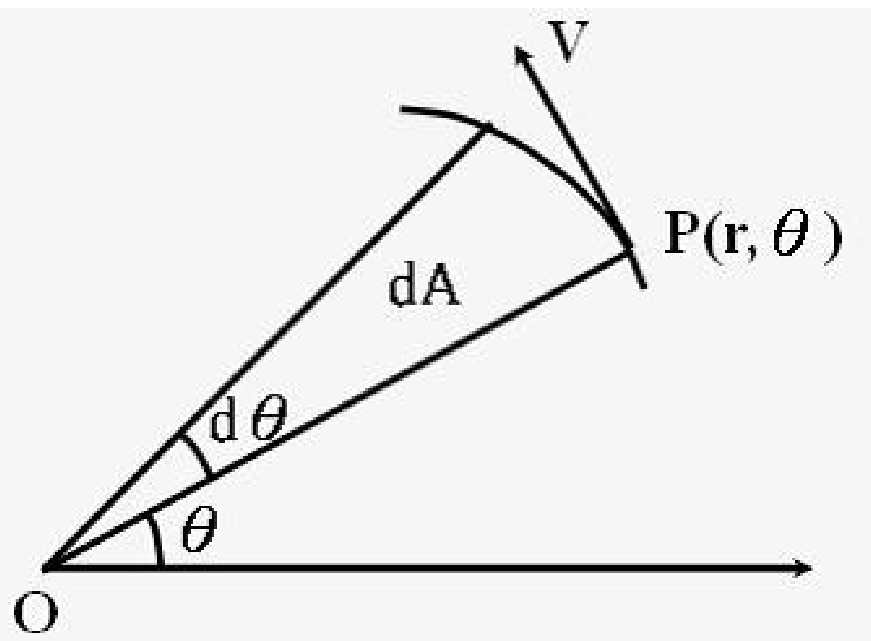}\\
 Figure 2
\end{center}

\end{enumerate}

\subsection{Mathematical analysis of Keplar's 1st and 2nd laws}

In this subsection, we shall present two alternative proofs of one of the
major result of Newton's Principia, namely.\newline
\emph{Theorem I}: Let $\mathbf{a}$ be the acceleration vector of a planet
motion as described by the 1st and the 2nd laws of Kepler. Then $\mathbf{a}$
is always pointing towards the sun and with its magnitude equal to $%
\displaystyle \frac{\pi ^2}{2}\frac{(2a)^3}{T^2}$ -times of the inverse of
the square of distance between the planet and the sun, namely
\begin{equation*}
| \mathbf{a} |=\frac{\pi ^2}{2}\frac{(2a)^3}{T^2}\frac{1}{r^2} \tag{$\ast$}
\end{equation*}
where $r$ is the distance, $T$ is the period and $2a$ is the length of the
major axis of its ellipse-orbit.\newline
\emph{Remark}: The direction of $\mathbf{a}$ is always pointing toward the
sun; this is exactly the kinematical significance of Kepler's 2nd law (cf. $%
\S $3.1). Thus, it suffices to prove the second assertion (i.e. ($\ast$)) on
the magnitude of $\mathbf{a}$.

\renewcommand{\proofname}{\textsl  First proof of ($\ast$):}
\begin{proof}

 As indicated in Figure 3, $\mathbf{v}$ (resp. $\mathbf{a}$) is the velocity (resp. acceleration) vector at P,
 $\{F'_1,F'_2\}$ are the reflectionally symmetric points of $\{F_1,F_2\}$ with respect to the tangent line $l_p$ and
 $\varepsilon = \angle M_1 P F_1$.\\
 Set $d_1$, $d_2$ to be the distances of $F_1$, $F_2$ toward $l_p$ and $h$ to
 be the height of $\triangle F_1 F'_1 F'_2$.\\

\begin{center}
 \includegraphics[scale=0.7]{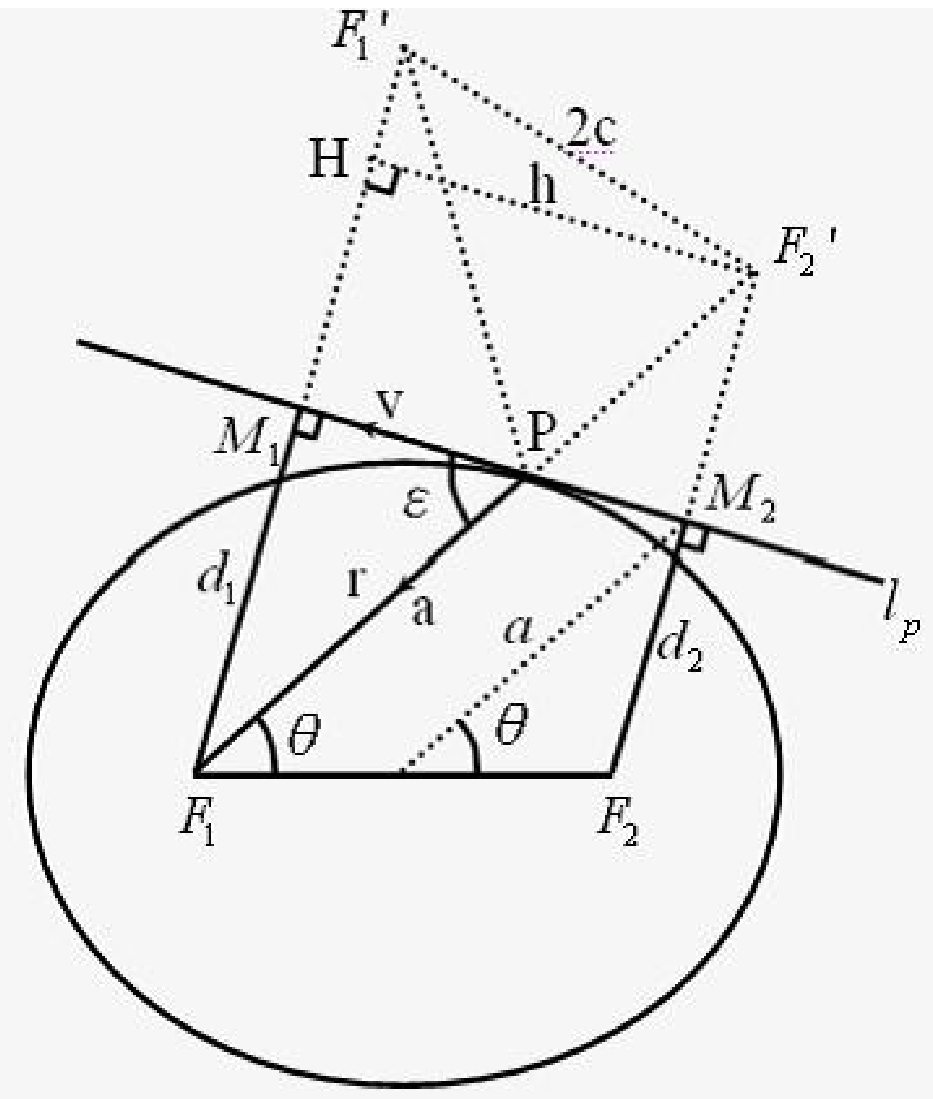}\\
 Figure 3
\end{center}

Then
 \[
 \begin{array}{l}
   4a^2 = \overline{F_1 F'_2}^2 = \overline{F_1 H}^2 + \overline{HF'_2}^2 =(d_1 +d_2)^2 +h^2 \\
   4c^2 = \overline{F'_1 F'_2}^2 = \overline{F'_1 H}^2 + \overline{HF'_2}^2 =(d_1 -d_2)^2 +h^2 \tag{10}
\end{array}
 \]
 Hence
 \[
 4b^2 =4a^2 -4c^2=(d_1 +d_2)^2 - (d_1 -d_2)^2 =4d_1 d_2 . \tag{11}
 \]
 On the other hand, Kepler's second law asserts that
 \[
 \begin{array}{lll}
  & |\mathbf{v}| r\sin \varepsilon = \frac{2 \pi ab}{T}, & r\sin \varepsilon=d_1 =\frac{b^2}{d_2}\\
  \Rightarrow & |\mathbf{v}|=\frac{2 \pi ab}{Td_1} =\frac{2 \pi a}{bT}\overline{F_2 M_2} \tag{12}
 \end{array}
 \]
 Note that
 \[
 \overrightarrow{F_2 M_2}= \overrightarrow{F_2 O}+\overrightarrow{OM_2}=
   \left(
   \begin{array}{c}
   -c\\
   0
   \end{array}
   \right)
   +
   a\left(
    \begin{array}{l}
    \cos \theta\\
    \sin \theta
    \end{array}
    \right)
   \tag{13}
 \]
 Therefore
 \[
 \mathbf{v}=
   \left(
   \begin{array}{cc}
   0&-1\\
   1&0
   \end{array}
   \right)
 \frac{2\pi a}{bT}\overrightarrow{F_2 M_2}=\frac{2\pi a}{bT}
   \left(
   \begin{array}{c}
   0\\
   -c
   \end{array}
   \right)
 +\frac{2\pi a^2}{bT}
   \left(
   \begin{array}{c}
   -\sin \theta\\
   \cos \theta
   \end{array}
   \right)
 \tag{14}
 \]
 and hence
 \[
 \begin{array}{cl}
 \mathbf{a}&= \frac{d}{dt}\mathbf{v}=\frac{d}{d\theta}\mathbf{v}\cdot \frac{d\theta}{dt}=\frac{2\pi a^2}{bT}
   \left(
   \begin{array}{c}
   -\cos \theta\\
   -\sin \theta
   \end{array}
   \right)
 \cdot \frac{2\pi ab}{T}\cdot \frac{1}{r^2}\\
 & =\frac{\pi ^2}{2}\cdot \frac{(2a)^3}{T^2}\cdot \frac{1}{r^2}
   \left(
   \begin{array}{c}
   -\cos \theta\\
   -\sin \theta
   \end{array}
   \right)
 \end{array}
 \tag{15}
 \]
\end{proof}

\emph{Remarks}:

\begin{enumerate}
\item[(i)] The above clean-cut simple proof also reveals the kinematical
meaning of the $3^{ \mbox{rd} }$ law.

\item[(ii)] In retrospect, the equation (11) already provides a "hand and
glove fitting" between the $2^{\mbox{nd}}$  law and the ellipticity. Thus,
it becomes very easy to deduce the simple formula (14) of $\mathbf{v}$,
from which the formula of $\mathbf{a}$ (i.e. (15)) follows immediately.
\end{enumerate}

\renewcommand{\proofname}{\textsl  Second proof of ($\ast$):}
\begin{proof}\emph{ }\\

\begin{center}
 \includegraphics[scale=0.5]{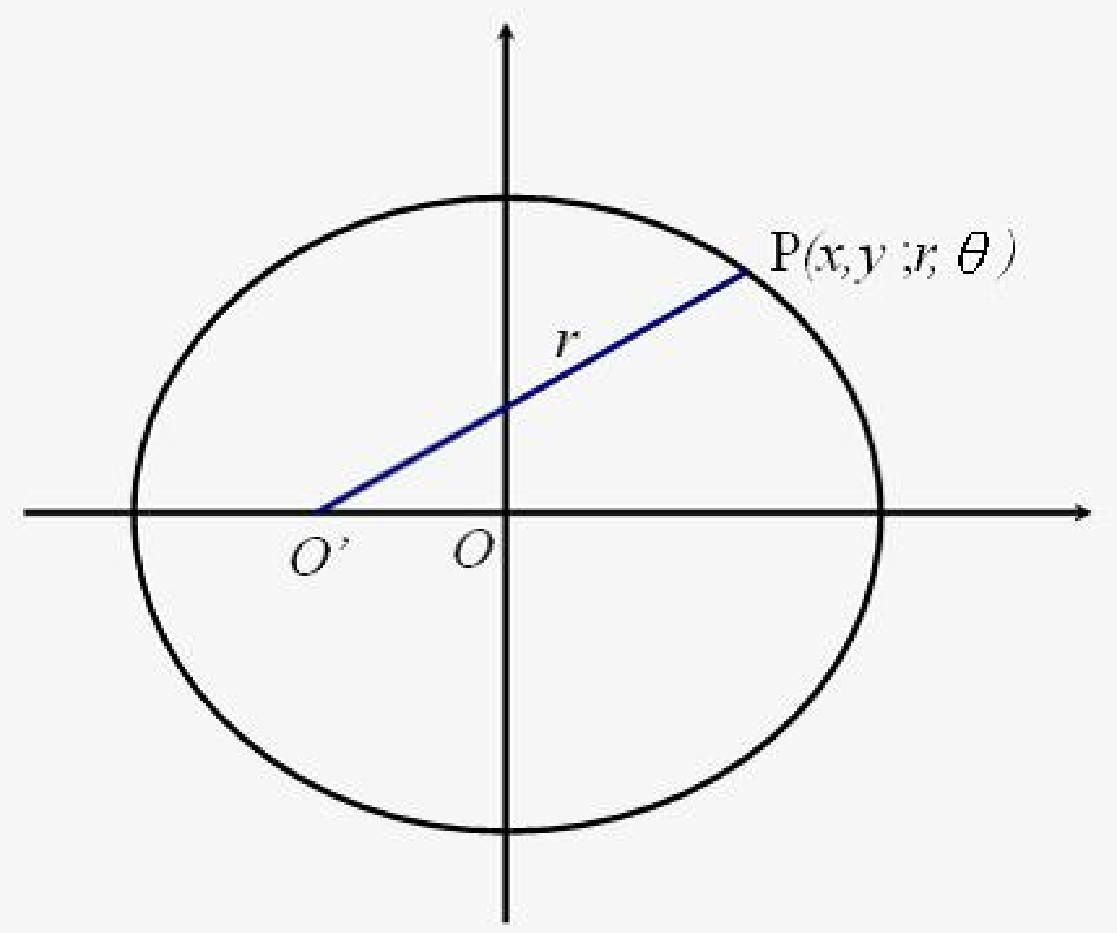}\\
 Figure 4
\end{center}

  As indicated in Figure 4, $(x,y)$(resp. $(r,\theta)$) are the Cartesian (resp. polar) coordinates of $P$
  with the origin $O$ (resp. $O'$) situated at its symmetric center (resp. one of its foci). Thus
  \[
  x=r\cos \theta -c, \: \: y=s\sin \theta \tag{16}
  \]
  and it is quite simple to deduce the following polar coordinate equation, namely
  \[
  \frac{1}{r}=\frac{a-c\cos \theta}{b^2} \tag{17}
  \]
  First of all, it follows from the $2^{\mbox{nd}}$ law that $\mathbf{a}$ is always pointing toward $O'$, namely
  \[
  \mathbf{a}=(\ddot{r} -r \omega ^2)
  \left(
  \begin{array}{c}
  \cos \theta\\
  \sin \theta
  \end{array}
  \right),
  \: \: \omega=\frac{d \theta}{dt} \: \: \mbox{(angular velocity)}
  \tag{18}
  \]
  Thus, it suffices to show that
  \[
  r^2 (\ddot{r} -r \omega ^2)=-\frac{\pi ^2}{2} \frac{(2a)^3}{T^2}
  \tag{$\ast'$}
  \]
  The following is a rather straightforward sequence of simple computations that deduces ($\ast'$)
  with the repeated help of the $2^{\mbox{nd}}$ law, namely
  \[
  \frac{dA}{dt}=\frac{1}{2}r^2 \omega= \frac{\mbox{total area}}{T} \: \: (=\frac{\pi ab}{T})
  \tag{$7'$}
  \]
  \begin{enumerate}
  \item [Step 1]: By the differentiation of (17) with respect to time, one has
        \[
        -\frac{\dot{r}}{r^2}=\frac{c}{b^2}\sin \theta\cdot \omega
        \tag{19}
        \]
        Thus
        \[
        \dot{r}=-\frac{c}{b^2}\sin \theta (r^2 \omega)=-\frac{2\pi ac}{bT}\sin \theta
        \tag{$19'$}
        \]
  \item [Step 2]: By the differentiation of ($19'$), one has
        \[
        \ddot{r}=-\frac{2\pi ac}{bT}\cos \theta \omega
        \tag{20}
        \]
        Hence, again using ($7'$)
        \[
        r^2 \ddot{r}=-\frac{2\pi ac}{bT}\cos \theta (r^2 \omega)=- \frac{4\pi ^2 a^2}{T^2}c \cos \theta
        \tag{21}
        \]
  \item [Step 3]: By ($7'$) and (17), one has
  \[
  \begin{array}{cl}
  r^2 (-r\omega ^2)&=-\frac{1}{r}(r^2 \omega)^2 =-\frac{1}{r}\frac{4\pi ^2 a^2 b^2}{T^2}\\
                   &=-\frac{4\pi ^2 a^3}{T^2}+\frac{4\pi ^2 a^2}{T^2}c\cos \theta
  \end{array}
  \tag{22}
  \]
  thus proving
  \[
  r^2 (\ddot{r}-r\omega ^2)=-\frac{\pi ^2}{2}\frac{(2a)^3}{T^2}
  \tag{$\ast'$}
  \]
  \end{enumerate}
\end{proof}

\emph{Remarks:}

\begin{enumerate}
\item[(i)] In comparison between the above two proofs, the first proof is
more geometrical, while the  second proof is more computational in nature;
and both of them are elementary, clean-cut and very simple.

\item[(ii)] In the second proof, one uses the $2^{\mbox{nd}}$ law four times
(namely, in obtaining (18), ($19^{\prime }$), (21)  and (22)) which enable
us to simplify the computations at each step , thus making the computations
altogether rather straightforward, elementary and simple. Of course one
still needs to differentiate  twice in order to compute the term of $r^2
\ddot{r}$ in ($\ast^{\prime }$). However, the above proof only uses the
analytical-geometric fact that
\begin{equation*}
\frac{d}{d\theta}  \left(
\begin{array}{c}
\cos \theta \\
\sin \theta%
\end{array}
\right)  =\left(
\begin{array}{c}
-\sin \theta \\
\cos \theta%
\end{array}
\right)
\end{equation*}
\end{enumerate}

\subsection{On the extension of Theorem I to the other two types of conic
sections}

Historically, Greek geometers first studied plane sections of right circular
cylinders, and discovered the remarkable characteristic property of such
curves (i.e. ellipses) of having a pair of foci with $\overline{F_1 P}+%
\overline{PF_2}$ equal to a constant. Later, they discovered that the
geometric proofs could be extended to prove similar results for plane
sections of right circular cones, which include two more type of curves,
namely hyperbola and parabola. Anyhow, it is again quite natural to seek
generalizations of Theorem I for centripetal motions with the other two
types of conic sections as their orbits, although such motions can hardly be
observed in celestial events simply because they are \emph{no longer}
periodic!\newline

\emph{Theorem $I^{\prime }$}: Let $\mathbf{a}$ be the acceleration vector of
a centripetal motion with the branch of hyperbola, as indicated in Figure 5,
as the orbit. Then
\begin{equation*}
\mathbf{a}=\frac{(2k)^2 a}{b^2}\cdot \frac{1}{r^2}  \left(
\begin{array}{c}
-\cos \theta \\
-\sin \theta%
\end{array}
\right) \tag{23}
\end{equation*}
where $2k$ is the constant value of $\displaystyle r^2 \frac{d\theta}{dt}$.

\renewcommand{\proofname}{\textsl  Proof:}
\begin{proof}
  As indicated in Figure 5, let \{$F'_i$\} be the reflection-image of $\{F_i\}$ and $d_i$ be the distances
  between $\{F_i ,F'_i \}$ and the tangent line $l_p$. Then
  \[
  \begin{array}{ll}
              &\overline{F_1 F_2}^2 =4c^2 =(d_1 +d_2)^2 +h^2, \: \: \: \overline{F_1 F'_2}^2 =4a^2 =(d_2 -d_1)^2 +h^2\\
  \Rightarrow &d_1 d_2=c^2 -a^2 =b^2
  \end{array}
  \tag{24}
  \]
  On the other hand, by the centripetality
  \[
  r^2 \frac{d\theta}{dt}=2 \frac{dA}{dt}=2k=|\mathbf{v}|r\sin \varepsilon =|\mathbf{v}|\cdot d_1
  \tag{25}
  \]
  Therefore
  \[
  |\mathbf{v}|=\frac{2k}{d_1}=\frac{2k}{b^2}d_2 =\frac{2k}{b^2}\overline{M_2 F_2}, \: \: \:
  \angle(\overrightarrow{M_2 F_2},\mathbf{v})=\frac{\pi}{2}
  \tag{26}
  \]
  Hence
  \[
  \mathbf{v}=\frac{2k}{b^2}
   \left(
   \begin{array}{cc}
   0&-1\\
   1&0
   \end{array}
   \right)
   (\overrightarrow{M_2 O}+\overrightarrow{OF_2})=\frac{2ka}{b^2}
   \left(
   \begin{array}{c}
   -\sin \theta\\
   \cos \theta
   \end{array}
   \right)
   +\frac{2k}{b^2}
   \left(
   \begin{array}{c}
   0\\
   c
   \end{array}
   \right)
   \tag{27}
  \]
  Thus having, by (25) and (27)
  \[
  \mathbf{a}=\frac{d\mathbf{v}}{d\theta}\frac{d\theta}{dt}=\frac{(2k)^2 a}{b^2}\frac{1}{r^2}
  \left(
  \begin{array}{c}
  -\cos \theta\\
  -\sin \theta
  \end{array}
  \right)
  \tag{28}
  \]
\end{proof}

\begin{center}
 \includegraphics[scale=0.7]{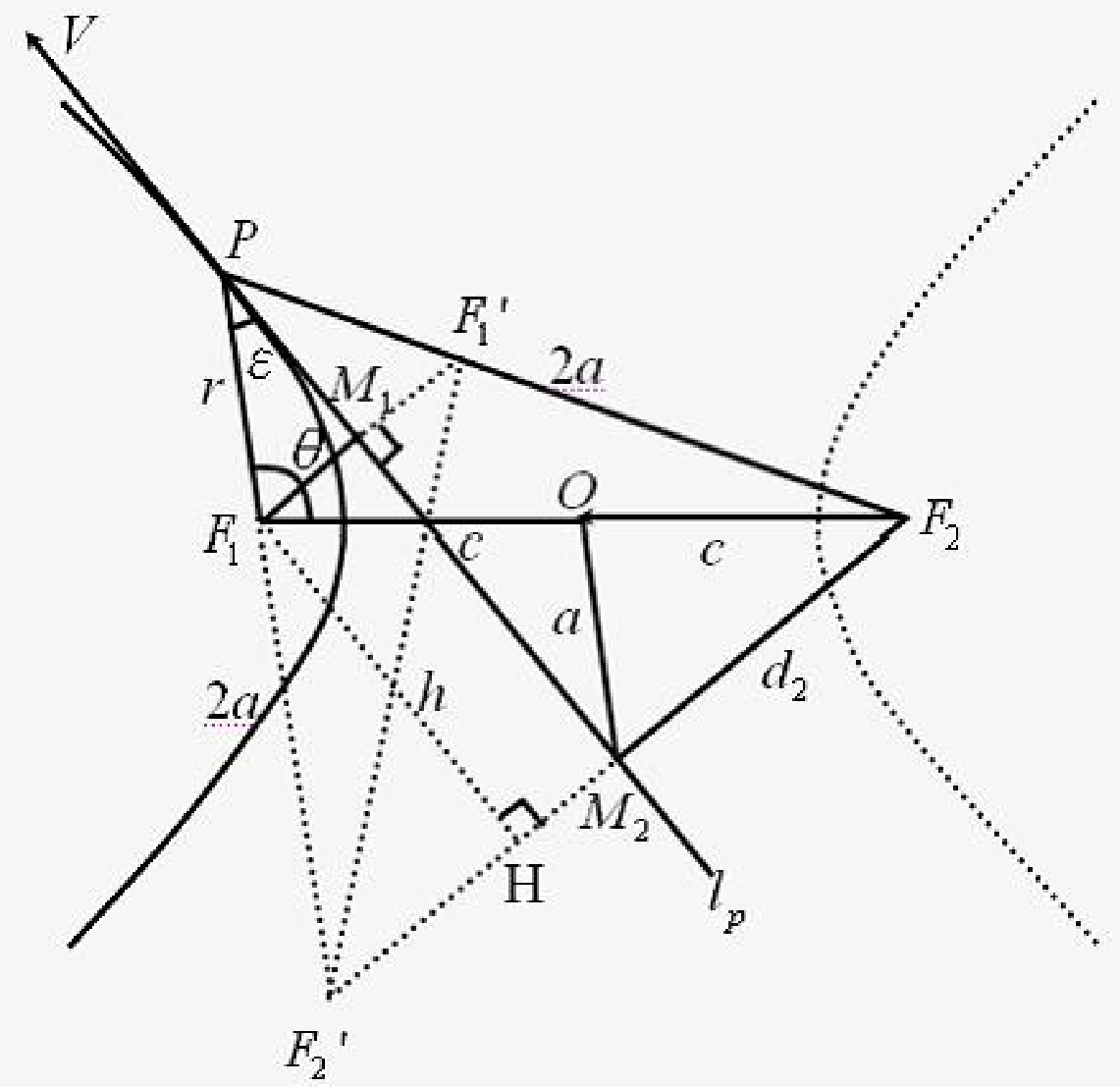}\\
 Fifure 5
\end{center}

\emph{Remark}: The same kind of proof will show that such a motion along the
other branch of hyperbola also satisfies the inverse square law, except it
will be \emph{repulsive} instead of attractive. In fact, such motions occur
naturally in the scattering theory [Fa].\newline
\emph{Theorem $I^{\prime \prime }$} : Let $\mathbf{a}$ be the acceleration
vector of a centripetal motion with the parabola, as indicated in Figure 6,
as its orbit. Then
\begin{equation*}
\mathbf{a}=\frac{2k^{2}}{p}\frac{1}{r^{2}}\left(
\begin{array}{c}
-\cos \theta  \\
-\sin \theta
\end{array}%
\right) \tag{29}
\end{equation*}

\begin{proof}
   By the centripetality, one has
   \[
   r^2 \frac{d\theta}{dt}=2\frac{dA}{dt}=2k=|\mathbf{v}|r\sin \varepsilon =|\mathbf{v}|\cdot d
   \]
   On the other hand, it is easy to see that
   \[
   \theta +\varepsilon =\theta +(\frac{\pi}{2}-\frac{\theta}{2})=\frac{\pi}{2}+\frac{\theta}{2}
   \]
   \[
   d=r\cos \frac{\theta}{2}, \: \: \: p=d\cos \frac{\theta}{2} =r\cos ^2 \frac{\theta}{2}
   \]
   Therefore,
   \[
   |\mathbf{v}|=\frac{2k}{r\cos \frac{\theta}{2}}=\frac{2k}{p}\cos \frac{\theta}{2}
   \tag{30}
   \]
   \[
   \mathbf{v}=\frac{2k}{p} \cos \frac{\theta}{2}
     \left(
     \begin{array}{c}
     -\sin \frac{\theta}{2}\\
     \cos \frac{\theta}{2}
     \end{array}
     \right)
   =\frac{k}{p}
     \left(
     \begin{array}{c}
     -\sin \theta\\
     \cos \theta
     \end{array}
     \right)
   + \left(
     \begin{array}{c}
     0\\
     \frac{k}{p}
     \end{array}
     \right)
   \tag{31}
   \]
   Hence
   \[
   \mathbf{a}=\frac{d\mathbf{v}}{d\theta}\cdot \frac{d\theta}{dt}=\frac{2k^2}{p}\frac{1}{r^2}
   \left(
   \begin{array}{c}
   -\cos \theta\\
   -\sin \theta
   \end{array}
   \right)
   \tag{32}
   \]
\end{proof}%

\begin{center}
 \includegraphics[scale=0.7]{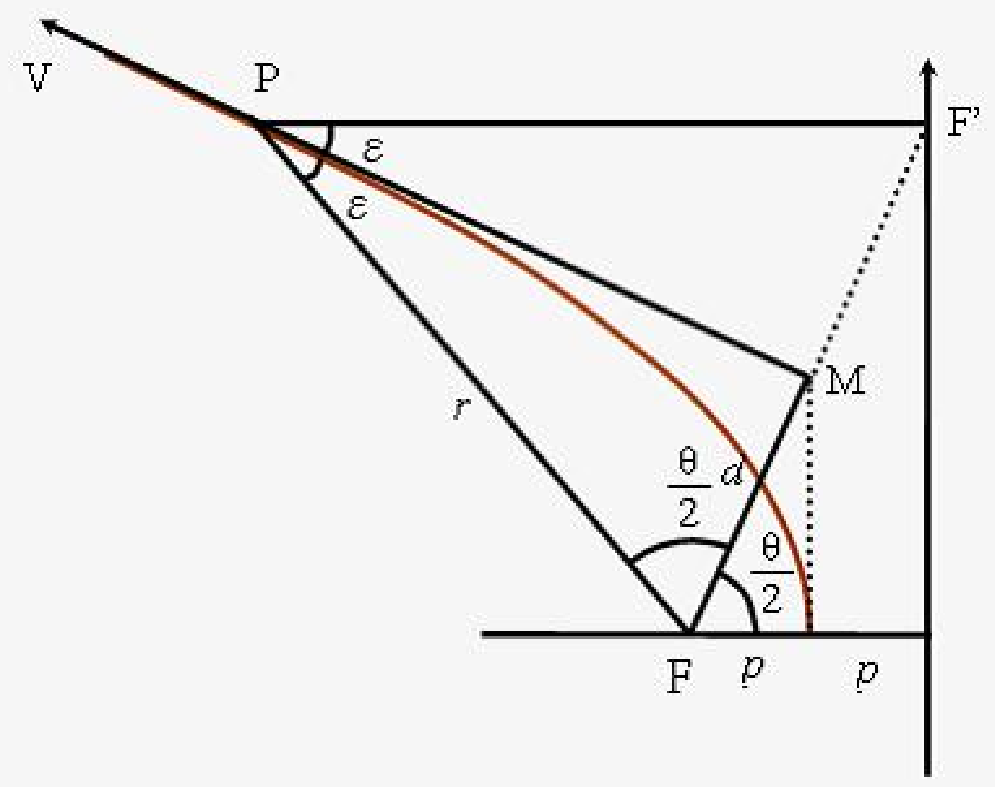}\\
 Figure 6
\end{center}

\emph{Remark}: By a simple coordinate transformation of $(r,\theta )$ to $(r,%
\tilde{\theta})$ with $\theta =\tilde{\theta}+\pi $, the equation (17) of
ellipse becomes
\begin{equation*}
\frac{1}{r}=\frac{a(1+e\cos \tilde{\theta})}{b^{2}}, \: \: \: ae=c\tag{$17'$}
\end{equation*}%
It is straightforward to check that, using the above polar coordinate
equation for hyperbola (resp. parabola), i.e. for the cases of $e>1$ (resp. $%
e=1$), the second proof of Theorem I automatically extend to that of Theorem
$I^{\prime }$ and Theorem $I^{\prime \prime }$, again by straightforward
differentiation and stepwise applications of $r^{2}\omega =2k$ (cf. \S %
(3.2)). Thus, the second proof of Theorem I actually also provides a proof
of both Theorem $I^{\prime }$ and Theorem $I^{\prime \prime }$ \emph{without}
modification, while the first proof of Theorem I can also be extended to
similar proofs of Theorem $I^{\prime }$ and Theorem $I^{\prime \prime }$
with some simple modifications, as above.

\subsection{The uniqueness theorem and the converse of Theorem I}

\emph{Theorem II}: Suppose that
\begin{equation*}
\mathbf{a}=\frac{K}{r^2}  \left(
\begin{array}{c}
-\cos \theta \\
-\sin \theta%
\end{array}
\right)  , \: \: \: K>0 \tag{$\ast '$}
\end{equation*}

Then the motion satisfies Kepler's second law and its orbit is a conic
section with the center as one of its faci.

\begin{proof}
  Centripetality implies that there exists a constant $k$ such that
  \[
    |\mathbf{v}|\cdot r \sin \varepsilon =2k, \: \: \: r^2 \frac{d\theta}{dt}=2k
    \tag{33}
  \]
  Therefore, it follows directly from ($\ast '$) and (33) that
  \[
  \frac{d}{d\theta} \mathbf{v}(\theta)=\mathbf{a}(\theta)\frac{dt}{d\theta}=\frac{K}{2k}\frac{d}{d\theta}
   \left(
   \begin{array}{c}
   -\sin \theta\\
   \cos \theta
   \end{array}
   \right)
  \tag{34}
  \]
  Hence, there exists a constant vector $\mathbf{c}$ such that
  \[
  \mathbf{v}(\theta)=\frac{K}{2k}
   \left(
   \begin{array}{c}
   -\sin \theta\\
   \cos \theta
   \end{array}
   \right)
  +\mathbf{c}
  \tag{35}
  \]
  Without loss of generality, we may assume that $r(\theta)$ is minimal at $\theta =0$ and $\mathbf{v}(0)$ is
  pointing upward, as indicated in Figure 7.\\

\begin{center}
 \includegraphics[scale=0.6]{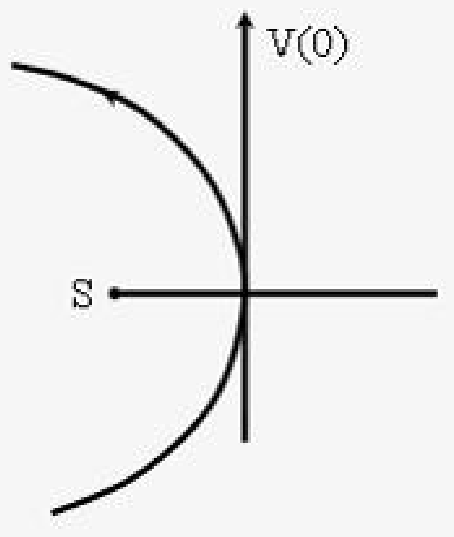}\\
 Figure 7
\end{center}

  Thus $\mathbf{c}$ is also pointing upward and
  \[
  \mathbf{v}(\theta)=\frac{K}{2k}
   \left(
   \begin{array}{c}
   -\sin \theta\\
   \cos \theta
   \end{array}
   \right)
  +
   \left(
   \begin{array}{c}
   0\\
   |\mathbf{c}|
   \end{array}
   \right)
  \tag{36}
  \]
  Now, again by the second law, we have
  \[
  r
   \left|
   \begin{array}{cc}
   \cos \theta & \frac{K}{2k}(-\sin \theta) \\
   \sin \theta & \frac{K}{2k}\cos \theta +|\mathbf{c}|
   \end{array}
   \right|
  =2k
  \tag{37}
  \]
  Namely
  \[
  \frac{1}{r}=\frac{K}{(2k)^2}(1+e\cos \theta), \: \: \: e=\frac{2k|\mathbf{c}|}{K}
  \tag{38}
  \]
  This is exactly the polar coordinate equation of a conic section with $e$ as its eccentricity!
\end{proof}

\emph{Remarks}:

\begin{enumerate}
\item[(i)] Note that it is again the second law that plays the important
role in the above very  simple straightforward proof. Conceptually, the
\emph{second law is actually the conservation law of angular momentum},which
is rooted in the rotational symmetry of the space.

\item[(ii)] In the case of elliptical orbit, one has (cf. ($7^{\prime }$))
\begin{equation*}
2k=\frac{2\pi ab}{T}, \: \: \: \frac{K}{2k}=\frac{2\pi a^2}{bT}, \: \: \: |%
\mathbf{c}|=\frac{2\pi ac}{bT}  \tag{39}
\end{equation*}
thus having
\begin{equation*}
\frac{K}{(2k)^2}=\frac{a}{b^2}, \: \: \: \frac{2k|\mathbf{c}|}{K}=e=\frac{c}{%
a}  \tag{40}
\end{equation*}
\end{enumerate}

\section{The gravitation force of a body with spherically symmetric density
exerting on an outside particle}

In this section, we shall present an alternative proof of the following
theorem which plays a fundamental, decisive r\^{o}le in Newton's discovery
of the law of universal gravitation (cf. Principia and the chapter 15 on
"the superb theorems" in [Ch].)\newline
\emph{Theorem III$'$}: The gravitation force of a body with spherically
symmetric density and total mass of M exerting on an outside particle of
mass m is equal to that of a particle of mass M situated at its center.%
\newline
\emph{Remark}: It is easy to see that the proof of Theorem III$'$ can be
reduced to that of the special case of a \emph{thin spherical shell with
uniform (area-wise) density}, namely.\newline
\emph{Theorem III}: Let $\Sigma $ be a thin spherical shell with uniform
(area-wise ) density $\rho $, radius R and P be an outside particle of mass
m. Then the total gravitation force of $\Sigma $ exerting on P is equal to
\begin{equation*}
G\frac{Mm}{\overline{OP}^{2}}, \: \: \: M=4\pi R^{2}\rho \tag{41}
\end{equation*}%
Where $G$ is the gravitation constant and $\overline{OP}$ is the distance
between the center of $\Sigma $ and $P$.

\begin{proof}
  As indicated in Figure 8, $\overline{OP}\cdot \overline{OP'}=R^2$. Therefore $\triangle OPQ$ and $\triangle OQP'$
  have a common angle at $O$ and
  \[
  \overline{OP} : \overline{OQ} =\overline{OQ} : \overline{OP'}
  \tag{42}
  \]
  thus having
  \[
  \begin{array}{cc}
    \triangle OPQ \sim \triangle OQP' ,& \\
    \angle OQP' =\angle OPQ \: (:=\theta), & \overline{P'Q} : \overline{QP} =\overline{OQ} : \overline{OP}
  \end{array}
  \tag{43}
  \]

\begin{center}
 \includegraphics[scale=0.7]{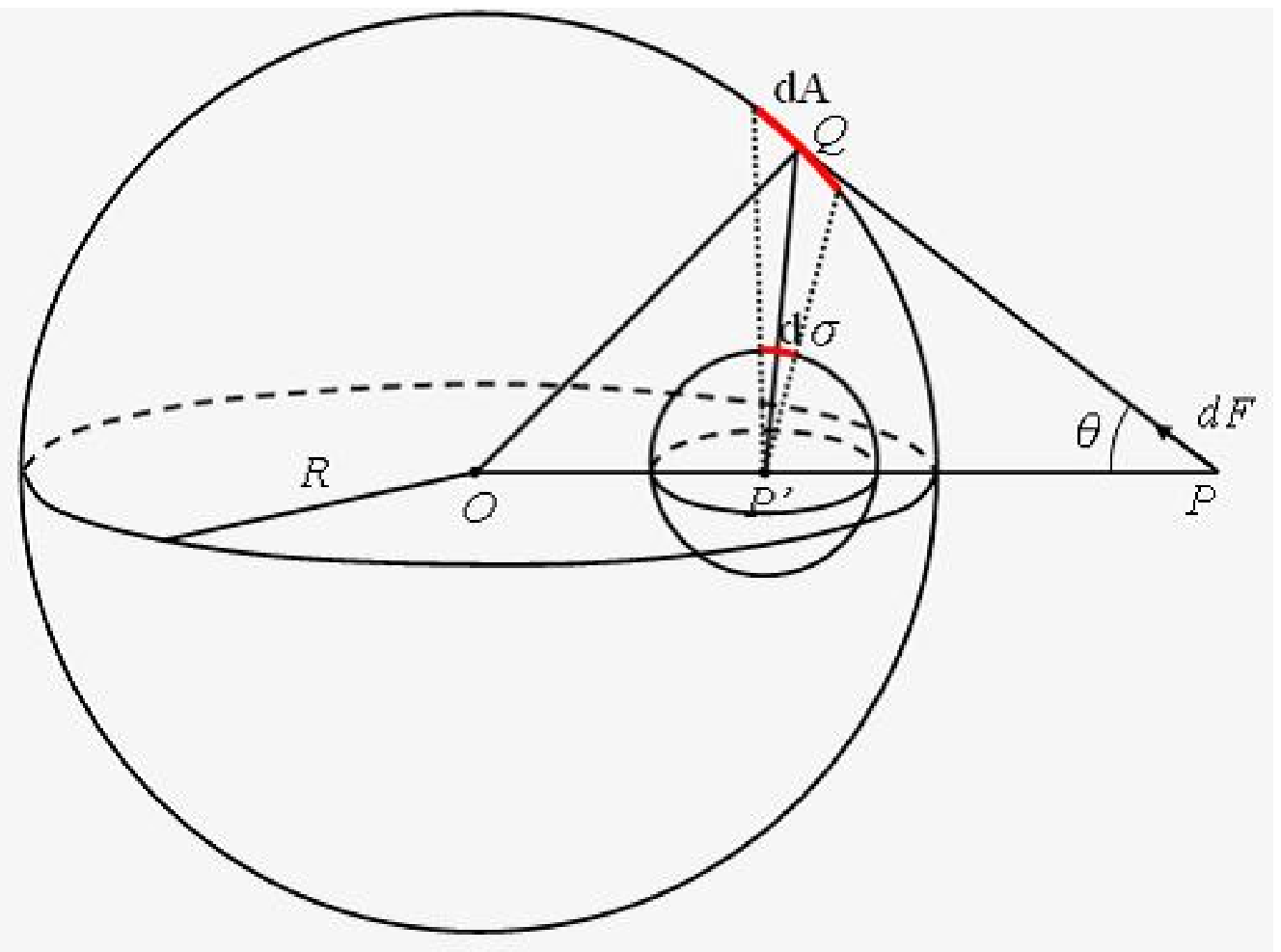}\\
 Figure 8
\end{center}

  Note that, it follows readily from the rotational symmetry of the whole geometric setting with respect to the
  line of $\overline{OP}$, the total gravitation force of $\Sigma$ exerting on $P$ is clearly in the direction
  of $\overrightarrow{PO}$. Thus, it suffices to find the total sum of
  \[
  |d\mathbf{F}|\cos \theta=G\frac{\rho dA\cdot m}{\overline{QP}^2}\cos \theta
  \tag{44}
  \]
  Set $d\sigma$ to be the solid angle of the cone with $dA$ as its base and $P'$ as its vertex. Then, as indicated
  in Figure 9
  \[
  dA\cos \theta =\overline{P'Q}^2 d\sigma .
  \tag{45}
  \]

\begin{center}
 \includegraphics[scale=0.5]{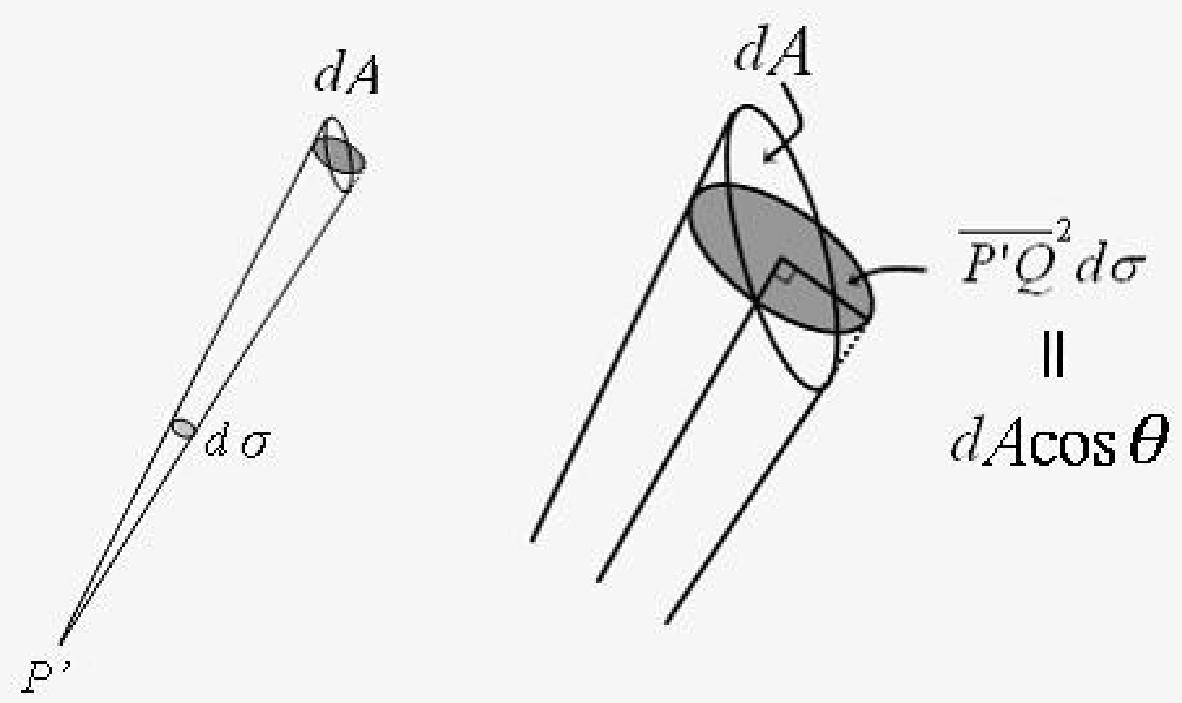}\\
 Figure 9
\end{center}

  Therefore
  \[
  |d\mathbf{F}|\cos \theta = Gm\rho \frac{\overline{P'Q}^2d\sigma}{\overline{QP}^2}=Gm\rho
  \frac{R^2}{\overline{OP}^2}d\sigma
  \tag{46}
  \]
  and hence, the total gravitation force is given by
  \[
  \sum Gm\rho \frac{R^2}{\overline{OP}^2}d\sigma=Gm\rho \frac{R^2}{\overline{OP}^2}\sum d\sigma
  =G\frac{4\pi R^2 \rho\cdot m}{\overline{OP}^2}=G\frac{Mm}{\overline{OP}^2}
  \tag{47}
  \]
\end{proof}

\emph{Remarks}:

\begin{enumerate}
\item[(i)] Note that a body with spherically symmetric density can be
regarded as the  \emph{non-overlapping union of concentric thin spherical
shells} with uniform area-wise  densities (depending on their radii). Thus
Theorem III$'$ is a direct corollary of Theorem III.  Moreover, the
same method of the above proof of Theorem III also shows that the total
gravitation  force between a pair of such bodies is, in fact, equal to
\begin{equation*}
G\frac{M_1 M_2}{\overline{O_1 O_2}^2}  \tag{48}
\end{equation*}
where $M_1$ and $M_2$ are their total masses and $\overline{O_1 O_2}$ is the
distance between their  centers (cf. $\S 5$ for a brief discussion on the
significance of such "superb theorems" [Ch].)

\item[(ii)] In retrospect, it is the advantageous subdivision of the
spherical surface $\Sigma$, corresponding  to the infinitesimal subdivision
of the total solid angle at $P^{\prime }$, that achieves the wonder of such
a remarkably simple proof. Geometrically speaking, only such a subdivision
can achieve the full  extend usage of the "sphericality " of the geometric
situation of the pair $\{ \Sigma , p \}$.
\end{enumerate}

\section{Grand synthesis and far-reaching generalization: Newton's law
of universal gravitation}

Historically, we may regard the Copernicus' book of 1543, "\emph{On the
revolutions of the Celestial Spheres}" as the grand opening salvo of the
\emph{scientific revolution}, while Newton's book of 1687, "\emph{%
Philosophiae Naturalis Principia Mathematic}a" was the triumphant
culmination of such a most important revolution. The crowning achievement
of Principia is the Newton's law of universal gravitation, based upon the
mathematical synthesis of Kepler's laws of planet-motion; and Galileo's
study of gravitation force and the basic principles of mechanics.\newline
As it has already been pointed out in the introduction, the major components that
enabled Newton to achieve the grand mathematical synthesis in Principia
consists of the four theorems as stated in $\S $1, whose proofs in Principia
(cf. [Ar] [Ch]) are elementary, geometrical but difficult to comprehend.
Now, with the simple, elementary and clean-cut proofs of $\S $3 and $\S $4
at hand, we hope that such simplifications will make the \emph{revisiting}
of Newton's epic journey from Kepler's and Galileo's laws to the universal
gravitation law enjoyable and inspiring for common reader, including earnest
high school students. Anyhow, the following are some additional crucial
ideas and highlights that may also be helpful for appreciating such a
journey:

\begin{enumerate}
\item[(1)] The discovery of Kepler's three laws on planet-motions is a
monumental milestone of the  civilization of rational mind which, for the
first time, grasp the wonderful organization of the solar system.  However,
they are empirical laws based upon the astronomical data of Tycho de Brahe
which are, themselves, only  with the assured accuracy of up to 2 minutes.
Moreover, they are only "verified" to be fitting for the six planets  of
Tycho's era; whether or not such laws still hold for other yet to be
discovered planets is another matter.

\item[(2)] Note that the sun and the planets are spherical bodies of huge
sizes just by themselves. However,  their sizes are comparatively much, much
smaller than the distances between them. Therefore, in the mathematical
analysis of the underlying reasons of Kepler's first and second laws (cf. $%
\S $3), it is still reasonable to regard  them as mere points. The proof of
Theorem I reveals that the "\emph{physical cause}" of planet-motions is a
kind  of attractive force toward the sun whose magnitude is inversely
proportionate to the square of distance. In  retrospect, one may regard the
elliptical orbits as a beautiful "\emph{hint}" of the Nature awaiting to
inspire  some mathematical mind to the discovery of inverse-square law.

\item[(3)] After realizing that heavenly motions of planets are, in fact,
governed by this kind of attractive  force, one nationally proceeds to
investigate whether the gravitation force studied by Galileo, or the force
that  keeps the moon circulating the earth are also the \emph{same kind of
force}? Here, one needs Theorem III (or III$'$)  in order to compute the total
gravitation force of the earth exerting on an earthy object or the moon.
Historically,  it is the difficulty of proving this theorem that delayed
Newton's publication of universal gravitation law.  (cf. [Ar] [Ch, p. 12, 13
and 302]).

\item[(4)] The importance of Theorem II lies in the vast applications of the
law of universal gravitation, rather than contributing to its \emph{discovery%
}. We refer to $\S $15 of [Ar] for a discussion whether Newton actually
proved Theorem II in Principia; and to [De] [Ha] [Ma] and [Go] for other
proofs of Theorems I and/or II.
\end{enumerate}

\end{document}